# *A critical analysis of models and experimental evidence of negative capacitance stabilization in a ferroelectric by capacitance matching to an adjacent dielectric layer*

J. A. Kittl[1,2], J. -P. Locquet[2], V. V. Afanas'ev[2]

[1]Advanced Logic Lab, Samsung Semiconductor Inc., Austin TX 78754, USA

[2]Department of Physics and Astronomy, KU Leuven, Leuven, Belgium

**Abstract**

We present a thorough analysis of the foundations of models of stabilization of negative capacitance (NC) in a ferroelectric (FE) layer by capacitance matching to a dielectric layer, which claim that the FE is stabilized in a low polarization state without FE polarization switching (non-switching), showing that the concept is fundamentally flawed and unphysical. We also analyze experimental evidence concluding that there is no data supporting the need to invoke such stabilization; rather, conventional models of ferroelectric polarization switching suffice to account for the effects observed. We analyze experimental evidence that at least in some of the model systems for which this effect has been claimed, categorically rule out stabilized non-switching NC. Microscopic measurements recently published as supporting non-switching stabilized NC actually rule them out, since the ferroelectric in a stack sandwiched between two dielectric layers was found to be in a mixed domain state (high polarizations within each domain) rather than in the low polarization state predicted by non-switching stabilized NC models. Nonetheless, since stabilized NC (corresponding to a minimum in free energy) is not physically impossible, it would be useful to move the research efforts to investigating scenarios and systems in which this effect is possible and expected and assess whether they are useful and practical for low power electronics

**Introduction**

During the past ten years, multiple papers have been published on advanced scaling of semiconductor devices under the premise that it is possible to stabilize a negative capacitance (NC) state in a ferroelectric (FE) layer simply by the effect of being adjacent to or connected (even if remote) to a dielectric (DE) layer, under a condition of "capacitance matching" [1-26]. These papers suggest that this stabilized negative capacitance (SNC) state in the ferroelectric corresponds to a local low polarization maximum of the free energy of the ferroelectric, and that the ferroelectric would adopt such an unfavorable local configuration simply because the adjacent (or remote) dielectric is in a very favorable low energy local configuration, as if the local configuration within the ferroelectric would be somehow determined ("how" is not explicitly described in these models) by the local configuration in a remote location within the dielectric.

The models and arguments are of the equilibrium (thermodynamic) type, claiming this low polarization negative capacitance state of the ferroelectric corresponds to the stable configuration of the system. As such, it should be possible to slightly perturb the system, and it would return to the same state. In addition, a quasi-static reversible trajectory of negative capacitance could be followed when slowly changing the external field. This trajectory, mapped in the polarization-electric field ($PE$) plane would have an S-shape (dotted red line in Fig. 1). Since, according to this models, all the FE would be in a low polarization state (rather than in a mixed domain state), simply oscillating around $P_{FE} = 0$ as the electric field changes, the response would be very fast (not requiring FE polarization switching, i.e. "non-switching"). If this was true, MOS transistors with gate stacks including DE and FE layers, or simply having a FE capacitor connected in series to a conventional gate, under the conditions of capacitance matching and operated at low fields (in region of SNC), could offer technological advantages, since higher charge in the transistor channel could be induced with smaller voltage swings. These devices would show a steeper sub-threshold slope (decades of transistor current change at given gate voltage change), SS, than those with simply thermal limited SS; i.e. these devices could achieve SS< 60 mV/dec, opening an exciting path for low power electronics.

NC requires a change in electric field in a material to induce a change of polarization in the opposite direction. The effect of electric fields on charge distributions typically results in the opposite effect, since positive charges are pushed along the field direction and negative charges in the opposite direction. While stabilization of a NC state in a material is not physically impossible, it is clear that this state would not occur at a local maximum of free energy (this would be unphysical). Additional forces (e.g. stress effects) would be needed for the local minimum of free energy (when all force fields are considered) that represent a physical stable state to correspond to a negative capacitance state. There have been experiments that show higher capacitance for stacks including ferroelectric layers than what would be obtained in a capacitor model if each layer would be considered separately [3]. This could be interpreted as an "effective" NC of the FE layer. In order to obtain devices with sub-60 mV/dec SS –the holy grail of low power electronics- through NC, it is necessary, however, for the whole gate insulator stack to have NC (not just high capacitance). To our knowledge, there is no experimental demonstration to date of a whole stack with overall stable NC. It is also important to note, that a non-linear positive capacitance element (e.g. normal FE behavior) can also lead to sub-60 mV/dec SS [27], however, this element would need to have no hysteresis for it to be useful for low power electronics.

On the other hand, specific dynamic conditions can lead to transient NC (TNC) during ferroelectric polarization switching [28, 29], which can result in NC (transient) of a whole stack, and can lead to sub-60 mV/dec SS. TNC is consistent with classical theories of ferroelectrics and happens when a change in the applied voltage triggers a relatively slow process involving a change in the atomic configuration of the material (ferroelectric polarization switching, a nucleation and growth type of process), and the voltage ramp direction is reversed (turning point) before the transition is completed. Once triggered, the transition continues in the same direction for some time (delay), while the voltage is already changing in the opposite direction, leading to transient NC [28, 29]. Existing experimental evidence of NC in ferroelectric layers can be explained by this dynamic, transient effect [28-32]. However, TNC is not appealing for technological applications in scaled low power transistors due to the typically low speed (fastest switching typically > 100 ps to ns) see Supplemental Materials), very limited operating conditions and limited endurance of FE polarization switching, and in most cases also due to hysteresis power loss.

Quasi-static NC in mixed domain systems has also been proposed [33-25]. A. M. Bratkovsky and P. Levanyuk proposed [33] that in FE capacitors with 180º domain structures and in the limit of thin dead layers, the overall capacitance increases significantly. They also point out that assuming a simple series capacitor model of the dead and FE layers, one would infer that the FE layer has negative capacitance, which they indicate to be an "artifact of the capacitor model" which they find to be "hardly applicable" to thin FE films and warn against the use of the capacitor analysis in this case ("clearly demonstrates the danger of applying a naïve electric circuit analysis to FE systems").

Recently, alternative theoretical models predicting quasistatic negative capacitance were proposed [34, 35]. In these models, the NC is attributed to effects of the depolarization field in specific mixed domain structures with a nanoscopic alternating domain structure, in which the response of the system is given by the motion of domain walls [34, 35]. There is a high contrast between these models, in which the FE is in a mixed domain state (high, FE polarization within each domain) and the models based on capacitance matching which predict the FE to be in a microscopic local low polarization state. We would classify these alternate models as "switching models" since by the passage of a domain wall, the local polarization undergoes normal FE polarization switching. It is argued that these mechanisms may be fast (no nucleation process needed, since the sample would remain always in the alternating mixed domain state), and could be also low loss (i.e. with minimal hysteresis) [35]. While these ideas remain an interesting topic of research, the conditions under which these type of domain structures are achieved in practice, the practical demonstration of these mechanisms, and fundamentally the applicability of these ideas to low power electronics remain unclear. Questions that would need to be answered include, which stacks would ensure the presence and stability of such domain structures, the possibility of integrating such stacks in modern electronic integrated circuits, including size and material constraints, the size of domains compared to the dimensions of devices, etc.). Also, as mentioned, for NC to result in sub-60 mV/dec (which would open the door for low power applications), the whole gate insulator built over the channel would have to exhibit NC (not just a high dielectric constant).

In this article, we discuss the basic physical foundations in which models of "non-switching" stabilized NC are based, use these physical foundations to point out the conceptual mistakes found in these models, and how this conceptual mistakes invalidate these models and their conclusions. We also analyze a wide range of experimental data showing that rather than supporting models of non-switching stabilized NC, they rather rule them out. The analysis presented here makes evident the fallacy of the concepts of non-switching SNC and in consequence, the lack of applicability of the types of devices inspired by these models to low power electronics.

**Physical foundations**

In physics, energy functions are defined such that their minimization under specific constraints give the equations of state (or constitutive relations) of the system (material). For example, the Helmholtz free energy $A(T, volume, x)$, where $T$ is the temperature and $x$ an internal configuration variable, is defined such that its minimization under constant temperature and volume results in the equation of state. If constraints are changed, e.g. we allow the volume to change, a Lagrange multiplier (in this case, the pressure ($P$)) is introduced: the function that is minimized to arrive to the correct equation of state under constant pressure and temperature is no longer $A$ but the Gibbs free energy $G(T,P,x)$ given by $G = A + PV$. Minimization of $A$ under constant $P$ and $T$ leads to erroneous equations of state.

The absolute minima of the appropriate energy function for the constraints considered give the stable states (from which the equation of state or constitutive properties are obtained), while local minima represent metastable states of the system. In both cases, small perturbations result in restoring driving forces that drive the system back to the minimum. For metastable states, kinetic considerations need to be included to determine the time at a given temperature that would take the system to move towards its thermodynamic equilibrium state. Local or global energy maxima are in contrast unstable states, i.e. small perturbations result in driving forces that lead the system away from the maxima. A maximum between a metastable and a stable equilibrium state provides a barrier for the system to move towards the stable (thermodynamic) equilibrium state, and as such may impact the kinetics of the transition to equilibrium, being this the only significance of a maximum in free energy.

A simple phenomenological model proposed by Landau [36] and later used by Devonshire [37] (See also Supplemental Materials), proposed, for a material close to a phase transition characterized by the vanishing of an order parameter (an internal variable, e.g. polarization for dielectrics and ferroelectrics), to approximate its free energy by a polynomial expansion on the order parameter. Symmetry considerations are used to decide which terms to keep in the expansion above and below the transition temperature.

We use the notation $\Phi_b\,(\boldsymbol{D},\boldsymbol{P},\sigma_{ij})$, $\widetilde{\Phi_b}(\boldsymbol{E},\boldsymbol{P},\sigma_{ij})$, and $\Phi_b^{\,*}(\boldsymbol{E}_{ext},\boldsymbol{P},\sigma_{ij})$ for the free energies at given stress $\sigma_{ij}$, with displacement, electric and external electric fields as independent variables (fixed in the minimizations) respectively.

Following the approach used in models of stabilized NC [1-5, 9, 15-19], we consider a one dimensional Landau model, with fields $E$ and polarizations $P$ along the z-direction; at fixed stress we have [36,37]:

$$\widetilde{\Phi_b}(E,P) = \frac{\alpha}{2}P^2 + \frac{\beta}{4}P^4 + \frac{\gamma}{6}P^6 - EP \qquad (4)$$

where $\alpha, \beta$ and $\gamma$ are temperature dependent material parameters, with $\alpha_{DE} > 0, \beta_{DE} = \gamma_{DE} = 0$ for linear dielectrics, $\alpha_{FE} < 0$ and $\beta_{FE}$ or $\gamma_{FE} > 0$ for a ferroelectric material in the FE state, as assumed in models of SNC (See also Supplemental Materials). In the simplest case of a linear dielectric we have: $\alpha_{DE} = 1/(\chi\epsilon_0)$ where $\chi$ is the susceptibility and $\epsilon_0$ is the permittivity of vacuum. Eq. (4) indicates the theoretical values of $\widetilde{\Phi_b}$ that the material would have if taking a specific polarization under a field. From these possible polarization values, only those corresponding to minima of $\widetilde{\Phi_b}$ are physically observed in equilibrium. For the DE, there is a single equilibrium polarization under an applied field, $P_{DE}(E) = \chi\epsilon_0 E$, which is the expected equation of state (or constitutive relation) for the DE, corresponding to the energy minimum $\widetilde{\Phi}_b^{DE} = -(\epsilon_0\chi/2)\,E^2$. We see how the correct energy construction results in the constitutive relation for the material by minimizing it with respect to the internal configuration variables ($P$) at fixed constraints (in this case $E$).

For FE, Fig. 1 illustrates the shape of the $\widetilde{\Phi}_b^{FE}$ energy surface. At a given (low) $E$, there are three extrema of $\widetilde{\Phi}_b^{FE}$ as a function of $P$, a global and a local minimum corresponding to stable and metastable equilibrium configurations respectively, and a local maximum corresponding to an unstable (thermodynamically meaningless) state. Microscopically, the minima typically correspond to two polar atomic configurations (shown in Fig. 1 for FE $HfO_2$ [38]), in which there are atomic displacements (either in +z or –z direction) from the non-polar configuration. The case of $P_{FE} = 0$ (non-polar) is microscopically unstable, and consequently not a physically possible equilibrium configuration. This state has meaning only as a kinetic barrier for the transition from the metastable to the stable state. At higher fields, there is only one global minimum. The extrema are projected in Fig.1 into the P-E plane where we show in black bold lines the stable configurations of the system, in thin blue lines the metastable configurations, and in dashed red lines the unstable (i.e. not equilibrium) configurations. Landau assigns absolutely no meaning to the dashed lines (which don't have any relation to equilibrium configurations) [36]. We note that the thermodynamically stable (i.e. in the limit of cycling rate going to 0) trajectory of the system is non-hysteretic. At $E = 0$, the system is broken into domains [36], some with positive polarization and some with negative polarization. Macroscopically, $P_{FE} = 0$ is achieved as an average over domains [36], or during polarization switching. In practical situations, at lab accessible cycling rates, the system remains in the same polarization state into the metastable region, and polarization switching occurs at $E \neq 0$; consequently, the $P - E$ trajectory exhibits hysteresis (Fig. 1, dotted green lines). A statistical analysis (e.g. following the Preisach model [28, 29, 39]) adequately describes the quasi-static hysteretic behavior of multi-domain FE.

The same analysis could be done alternatively, using the external field as independent variable, which is common in the analysis of FE [40, 41]:

$$\Phi_b^*(E_{ext}, P) = \frac{\alpha^*}{2}P^2 + \frac{\beta^*}{4}P^4 + \frac{\gamma^*}{6}P^6 - E_{ext}P \quad (5)$$

Where $\alpha_{FE}^* < 0$ and $\beta_{FE}^*$ or $\gamma_{FE}^* > 0$. The material parameters are adjusted to result by minimization in the same equations of state as derived from eq. 5 (otherwise, the formalism would be useless to describe a FE).

This formalism makes analysis of bilayer or multilayer capacitors straightforward, since the external field is the same in all layers.

We note that rarely some authors have used free energies with $E$ and $D$ as independent variables [42], and even mimic Landau's expression:

$$\psi(D) = \frac{\alpha'}{2}D^2 + \frac{\beta'}{4}D^4 + \frac{\gamma'}{6}D^6 \quad (6)$$

With the equation of state given by:

$$E = \alpha'D + \beta'D^3 + \gamma'D^5 \quad (7)$$

It is simple to show, by expanding the equations of state obtained from eqs. (6) and (7) (using $D = \varepsilon_0 E + P$), that they are completely incompatible with the equations of state obtained from Landau's formalism (eq. (4)), i.e. if Landau's equation (4) represent correctly a FE (as clearly established in literature), then eqs. (6) and (7) do not (see also Supplemental Materials).

Lastly, we note that Landau's is a mean field theory and should not be applied to regions in which parameters (e.g. fields, polarization and stress) are varying strongly in microscopic distances. Models including at least gradient terms (that may dominate the free energy behavior) should be used instead in these situations.

### Models of stabilization of a FE in a non-switching negative capacitance state by the presence of an adjacent dielectric

A series of models have been proposed in the past ten years that claim that a FE can be stabilized in a state of negative capacitance (which in these models corresponds to the red dashed line of unstable configurations in Fig. 1), corresponding to a local low polarization state in the FE and without FE polarization switching, simply by the presence of an adjacent dielectric under a capacitance matching condition [1-5, 9, 16-24]. The configurations discussed include DE-FE bilayers or multilayers and even separate DE and FE capacitors connected in series (i.e. with at least one metal plate in between). There is a persistent insistence on trying to claim or prove that indeed the FE is stabilized in a local configuration at a maximum of free energy, simply because the DE is (remotely) in a stable configuration at a (deeper) minimum in free energy [1-5, 9, 16-23].

The key trick used in several of these models is to plot the free energies of the FE and DE layers on a single axis, i.e. vs. the same independent variable so that they can be added at each value of the axis (variable), as shown in Fig. 2. The Landau expressions (eqs. (4) or (5)) are used for the free energy of the FE and DE layers (with appropriate material parameters to represent FE and DE respectively). There are to our knowledge at least three versions of this analysis. Two of these take Landau's form given in eqs. (4) or (5) and simply replace the internal configuration variable $P$ by either $Q_f$ (free charge in capacitor plates) [1, 5-9, 11, 15-23] or $D$ [26] (displacement field, which is a proxy for charge in the capacitor plates for the one-dimensional system, i.e. fixed $D \Leftrightarrow$ fixed charge in the plates of the capacitor) [26]. Needless to say, the physics of the systems described by these modified equations are in no way related to ferroelectrics (or any other material to our knowledge). If we do consider FE and DE materials, then we can apply Landau's equations with the internal configuration variable, $P$, as the independent variable [36, 37], as described in detail by Landau et al. [36], Devonshire et al. [37] and most authors since then. Here, we run into an obvious problem: the local polarizations in the DE and in the FE layers can indeed be different. There is no common axis to plot and add the free energies. In order to continue the stabilized negative capacitance argument, it is necessary to assume that the polarizations in the DE and FE layers are the same. This was recognized in some of the papers on stabilized NC [3], and analyzed in detail in [32]. This assumption is unphysical for separate capacitors. We argued in [32] that adding a polarization coupling term (proportional to $(P_{DE} - P_{FE})^2$) to the system energy for bilayer systems could indeed result in this condition being satisfied, and that under some conditions, the DE-FE stack may behave as a dielectric with higher capacitance than just the DE layer alone. However, we clarified that in this case, the FE is *not* in a negative capacitance state. Rather, since the polarization is similar or equal in the two layers, both layers behave as higher permittivity dielectrics, i.e. the electrical response of the two layers is now identical and equivalent to that of a dielectric.

A second line of reasoning has also been used to argue in favor of NC stabilization in FE-DE systems [23] (See also Supplemental Materials). This argument applies equally to separate capacitors or bilayers. The free energies $\widetilde{\Phi}_b(E, P)$ are used for the FE and DE layers. A condition of fixed total voltage at the capacitor plates for a bilayer, or fixed total voltage from the top of one capacitor to the bottom of the other capacitor, is used as constraint, and the addition of the free energies of the DE and FE layers ($\widetilde{\Phi}_b^{FE}(E^{FE}, P^{FE}) + \widetilde{\Phi}_b^{DE}(E^{DE}, P^{DE})$) is minimized under this constraint. It is clear that this condition is not equivalent to fixed fields $E^{FE}$ and $E^{DE}$ in each layer. As explained in detail in the section on physical foundations, the minimization of free energies leads to the correct equations of state *only* when the appropriate independent variables are kept fixed during the minimization. The results of minimization of $\widetilde{\Phi}_b^{FE}(E^{FE}, P^{FE}) + \widetilde{\Phi}_b^{DE}(E^{DE}, P^{DE})$ under fixed total voltage (across both layers) do not represent physical situations. As mentioned in the physical foundations section, the free energy $\Phi_b^*(E_{ext}, P)$ is much more convenient for the analysis of bilayer or multilayer capacitor structures, since (assuming no free charges in the layers or between the layers) at fixed external free charges in the plates of the planar capacitors, we have $E_{ext}$ constant across the structure. The analysis presented in [23] supporting NC stabilization in a FE by capacitance matching to a DE breaks down completely when using correctly the formalism with $\Phi_b^*(E_{ext}, P)$.

**Experimental evidence**

Many experiments have been quoted as supporting the models of stabilization of non-switching NC in a FE by a DE [3, 4, 6-13, 24-26].

One set of experiments use MOS transistors with either a DE-FE gate insulator stack or a standard DE insulator in the gate connected in series to a FE capacitor [4, 6, 9-14, 24, 25]. Observation of sub-60 mV/dec SS has been quoted as a proof of NC stabilization in the FE [6-10, 12, 13]. However, the sub-60 mV/dec SS can be easily explained through conventional FE switching (which basically changes the threshold voltage of the transistor), and does not require invoking arguments of NC stabilization [28-32, 43, 44]. Small or apparently lack of hysteresis in some experiments can be explained as a cancellation between counter-clockwise ferroelectric hysteresis and clockwise charge trapping-detrapping hysteresis (a well understood phenomena in MOS devices) [28, 29, 32], as illustrated in Fig. 3. In addition, NC during ferroelectric polarization switching might lead under specific cycling conditions to negligible hysteresis [35].

A recent set of experiments can be used to categorically rule out non-switching stabilized NC in specific DE-FE systems [24]. Small signal capacitance measurements of DE-FE stacks for varying DE thicknesses (to ensure capacitance matching would be met at some thickness) showed in all cases the total capacitance to be equivalent to the series capacitance of the two capacitors (for small signals, no FE P switching occurs, and the FE stays in the same P state) (Fig. 4). If stabilized NC was correct, then the NC trajectory would be stable and reversible (an equilibrium trajectory) and should be observed under a wide range of experimental conditions, including specifically these small signal measurements. A similar experiment for the case of a FE capacitor connected in series with DE capacitors also ruled out categorically the possibility of stabilized NC in these systems [45].

The same FE-DE stack, for which stabilized NC was ruled out in [24], was used in the same work to "build" an S-curve like plot [24]. Similar results were presented in [25]. Pulsed measurements were performed with pulses in the μs timescales increasing the maximum pulse voltage gradually and sequentially [24, 25]. The samples had been first taken to a high negative voltage to ensure orientation of domains in the opposite direction. Pulse voltage was gradually increased until voltages were reached for which domains within the FE layer would start to switch polarization. While minor loop Q-V plots (actual experimental trajectory) were not shown, a Q-V plot consisting of extracted parameters with each point corresponding to a calculation performed for a different pulse was presented. This plot extracted through manipulation of data from a highly dynamic experiment involving FE switching was presented as a measurement of the S-curve, which is supposed to be a thermodynamically stable trajectory, and postulated to give access to the shape of the free energy curve of the FE. While the energy barrier between the metastable and stable polarization states may have only an implication on the kinetics of the transition (which are actually more complex involving typically nucleation and growth of domains), there was no modelling of the dynamics, and thus, no information on the barrier should be concluded. While the NC effects were recognized in [24, 25] to be transient (TNC) and not stabilized, it was suggested that these could still be useful due to lack of hysteresis. We obtained the data from the experiments in ref. [25] from the authors. Q-V minor loop plots for the FE-DE capacitor (Fig. 4) show clear and conventional hysteretic behavior consistent with what is expected for FE switching.

A few experiments using epitaxial perovskite DE-FE structures have shown an enhancement of capacitance of the stack compared to the capacitance of the DE layer alone [3]. As mentioned before, for coupled systems, there is no reason why the system properties should be a simple combination of the properties of the layers. Thus, these experiments don't prove stabilized NC in the FE; rather only an "apparent" NC [32]. Strong polarization coupling, postulated for these systems [46, 47], could explain these findings [32]; if this was the mechanism at play then there wouldn't be a true NC effect and it would not be possible to produce stacks with true NC (rather than overall higher permittivity) which are needed to achieve sub-60 mV/dec SS through NC.

Recently, a quite different set of experiments was presented as supporting models of stabilized NC in a FE by capacitance matching to an adjacent DE [26]. In [26], a cross sectional transmission electron microscopy (TEM) study of DE-FE-DE epitaxial perovskite stacks provided two dimensional maps of polarization and electric field in the stack. The first striking observation is that the FE layer is not in a low polarization state as predicted by "non-switching" stabilized NC models, rather it is in the conventional mixed domain state. This observation alone suffices to rule out the applicability of "non-switching" stabilized NC models to the stack studied.

Several approximations are used in [26] to attempt to extract the variation of a local free energy of the FE across a domain boundary at the middle of the FE layer (away from the DE interfaces), in a direction parallel to the DE-FE interface. We note that along the path considered (across a domain wall), *all* parameters (including polarization, fields and strain) have large microscopic variations. In the analysis, $D$ is used as independent variable. While it is correct to assume that the *partial* derivatives of appropriate free energy functions w.r.t. the components of $\boldsymbol{D}$ give the respective components of $\boldsymbol{E}$ (e.g. $\frac{\partial \Phi_b}{\partial D_z}|_{Dy,Dz,\boldsymbol{P},\sigma_{ij}} = E_z$) [36], it is mathematically and significantly incorrect to assume that the variation of these functions along paths in which *all* variables are varying strongly should be equal to the electric field. Furthermore, across a domain wall boundary, gradient terms likely dominate the free energy behavior. It is not unexpected, in any case, that the free energy of a domain wall would be higher than the free energy at the center of a domain (otherwise, the sample would generate spontaneously a huge amount of domain walls), in the same way that the free energy of a grain boundary is expected to be higher than within a grain in polycrystalline samples. Domain walls are accidents of specific kinetic paths (e.g. in nucleation and growth processes), and their higher energy a consequence of their constraints (typically two *different* low energy configurations at each side). Let's indeed assume that the free energy moving along a path normal to the domain wall has a maximum at the domain wall. This does not imply that the domain wall is, given its constraints (two opposite polarizations at each side and close to it), locally at its maximum of free energy, that would be unphysical. There seems to be an obsession to prove that a local configuration which corresponds to a maximum of free energy (physically impossible) can be favored as a stable state. Instead, we suggest the domain wall chooses its internal configuration to minimize its local free energy *given* the external constraints. In any case, rather than supporting models of non-switching stabilized NC, ref. [26] would attribute the negative capacitance to the motion or passage (induced by electric fields) of 180º domain walls, which could be a path for low power electronics.

**Conclusions**

In conclusion, we have shown that there is no theoretical nor experimental validation of models of stabilization of non-switching NC in a FE by an adjacent DE layer through "capacitance matching". Furthermore, we have shown that experimental evidence from electrical data, at least in some of the model systems for which this effect has been claimed, actually categorically rule out the possibility of this stabilized NC. Furthermore, we highlight that microscopic measurements recently published as supporting stabilized NC models actually rule them out, since the ferroelectric in a stack sandwiched between two dielectric layers, was clearly found to be in a mixed domain state (with high polarizations within each domain) rather than in the low polarization state predicted by stabilized non-switching NC models. Since stable NC (corresponding to a minimum in free energy) is not physically impossible, it would be useful to investigate instead scenarios and systems in which this effect is actually expected.

**Figures:**

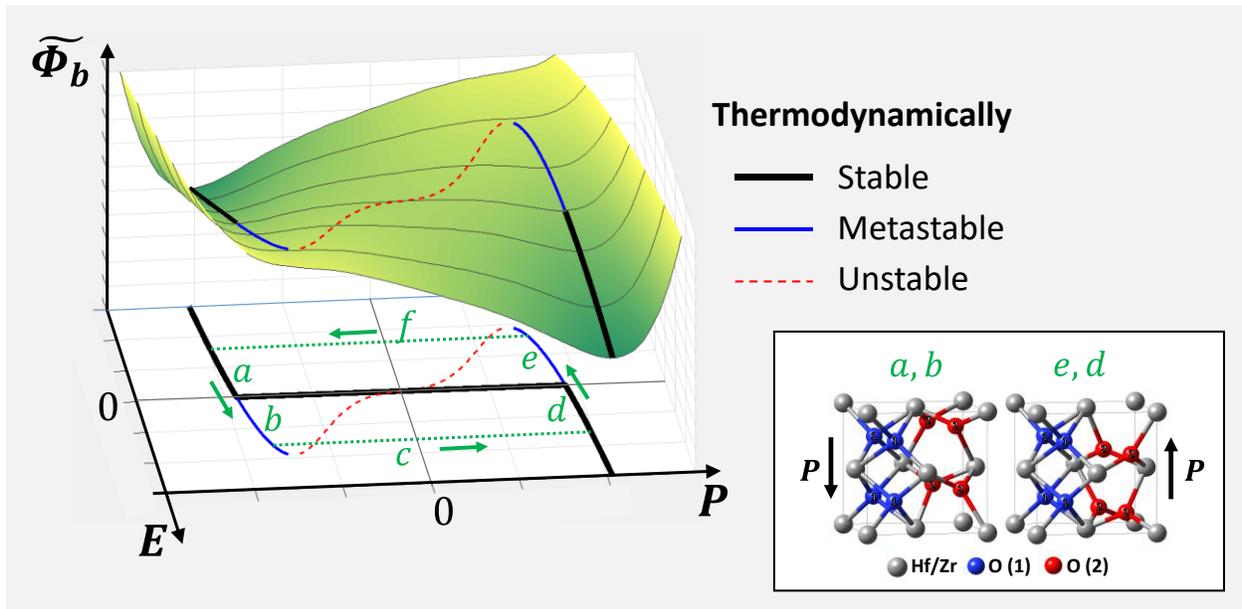

**Fig. 1**.: Plot of free energy $\widetilde{\Phi}_b(E, P)$ of a ferroelectric using Landau equation for the 1D case ($E$ and $P$: components of the electric field $\boldsymbol{E}$ and polarization $\boldsymbol{P}$ along the z-direction). At small fields, there are three extrema w.r.t. polarization: the global minimum corresponding to the thermodynamically stable state (bold black lines), a local minimum corresponding to a thermodynamically metastable state (blue lines), and the maximum corresponding to an unstable state (dashed red lines) which is not meaningful in thermodynamics. At higher fields, there is only a global minimum. When cycling the ferroelectric at limiting slow rates, the reversible bold black line trajectory is followed, which corresponds at $E = 0$ to a mixed domain state (locally $P$ is either up or down, not 0; this is similar to tie lines corresponding to mix-phase states in binary-alloy phase diagrams at fixed temperatures). For practical cycling rates, the metastable states are accessed, and the trajectory (schematically represented in dotted green lines) goes from a stable state ($a$ or $d$) to a metastable state ($b$ or $e$) before polarization switching ($c$ or $f$) into a stable state ($d$ or $a$). The field at which the polarization switching occurs (coercive field) varies with cycling rate, since the transition from the metastable to the stable state is controlled by the kinetics of polarization reversal. Statistical (e.g. Preisach) models can be used for ferroelectrics capturing the distribution of coercive fields (at a given cycling rate) in multi-domain samples. Stabilized negative capacitance models suggest the red dashed line trajectory ("S-curve") corresponding to the maximum of free energy can be stabilized into a reversible thermodynamic path. The inset shows as example the structures of ferroelectric Hf-Zr oxides for down (left) and up (right) polarizations. Ferroelectrics have polar structures (e.g. in the schematics, a clear distinction between up and down directions) with atoms displaced from the non-polar configurations (for which up and down directions are equivalent) which are unstable. Thermodynamic equilibrium requires the configuration to be stable or metastable, thus the non-polar configurations are not observed other than during P switching.

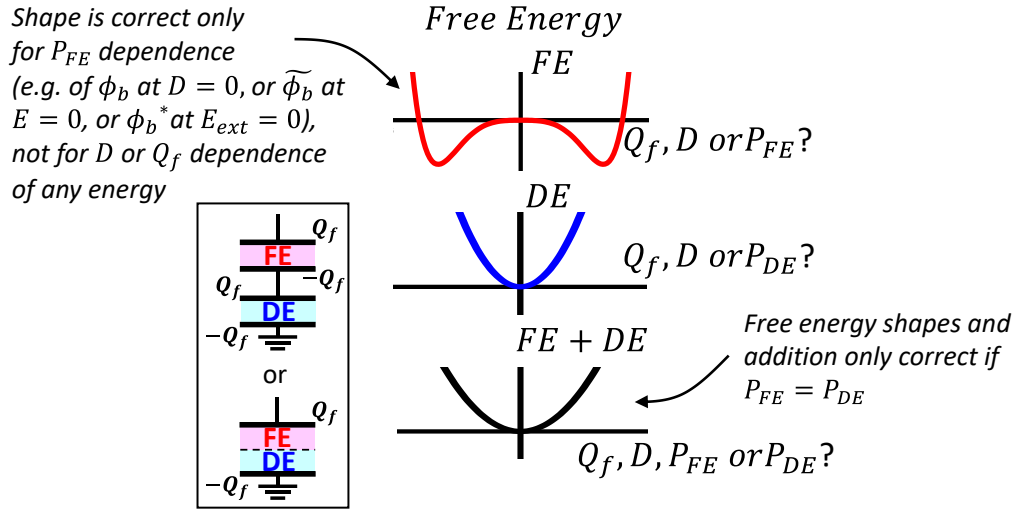

**Fig. 2**: Schematic of stabilized negative capacitance ansatz. Assumes the free energy shapes of the ferroelectric and dielectric layers as shown in the schematic, and that they can be added point by point as a function of a common variable for the two layers. The shape of the FE free energy is only correct for the dependence on the polarization of the FE ($P_{FE}$). In order to add point by point the free energies of the DE and FE layers, using the shapes shown, the common axis would then be a polarization, implying that $P_{FE} = P_{DE}$. We have shown [32] that if strong polarization coupling between the layers is present, then this would be satisfied. However, in the coupled system, the two layers would have the same polarization, and consequently the same electrical behavior, thus the two layers would behave as higher dielectric constant dielectrics (with the same dielectric constant) and there would be no layer with negative capacitance. Note that in coupled systems, the properties of the system are in general not a simple combination of the properties of the two layers.

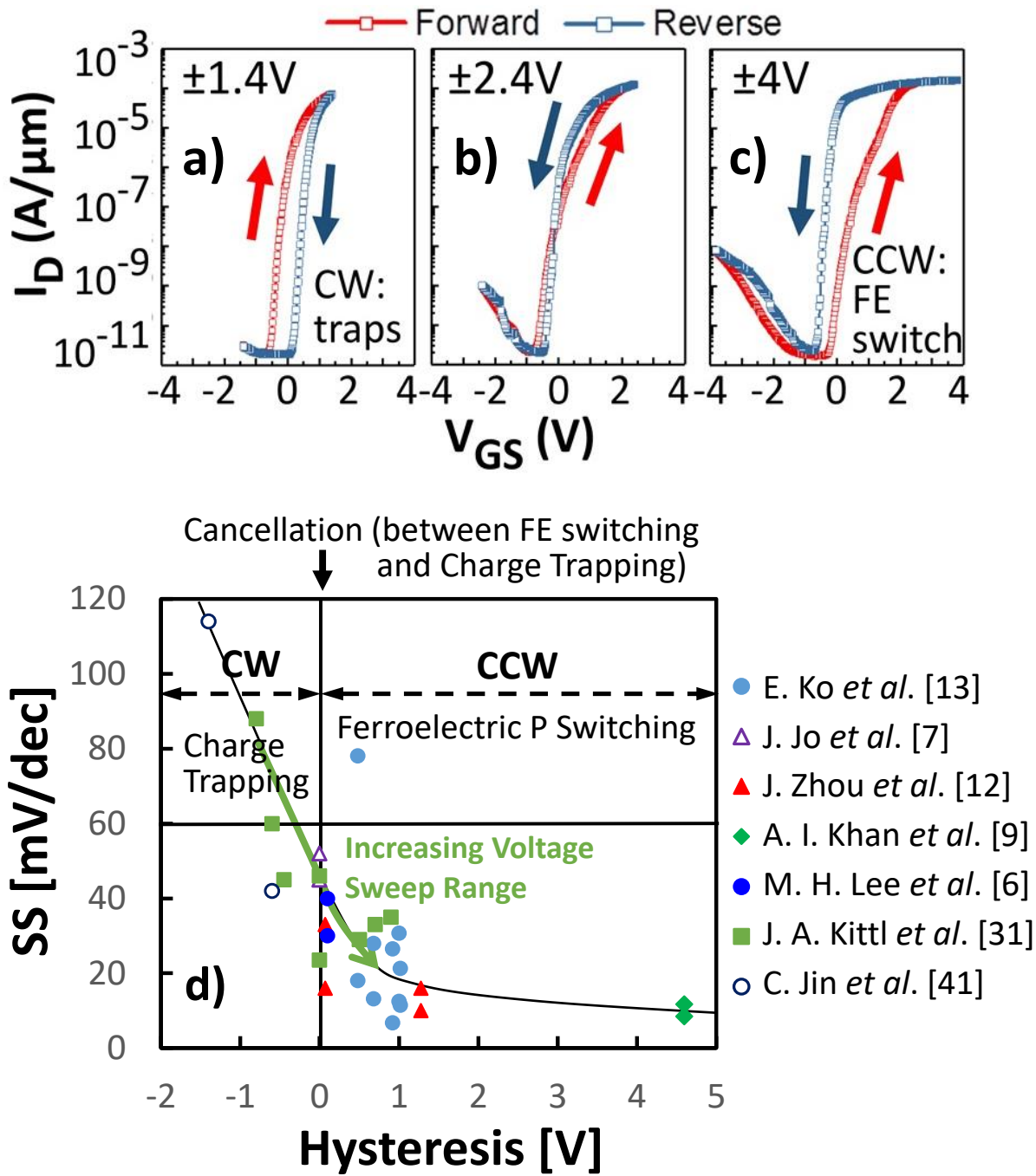

**Fig. 3**: Negligible hysteresis in some experiments can be attributed to cancellation between FE hysteresis and charge trapping-detrapping hysteresis. (a, b, c) Drive current ($I_D$) vs gate voltage ($V_{GS}$) for MOS transistors with a FE layer in the gate stack, for different gate voltage $V_{GS}$ sweep ranges. At low sweep voltage ranges (a) hysteresis is dominated by trapping-detrapping (clockwise (CW)) since there is no significant FE $P$ switching. At high voltage sweep ranges (c), hysteresis is dominated by ferroelectric $P$ switching (counter-clockwise (CCW)). At intermediate voltage sweep ranges, the superposition of the two mechanisms results in low hysteresis, however ferroelectric $P$ switching losses are still present. This compensation is present only for limited experimental conditions. (reproduced from ref. [32]). (d) Sub-threshold slope vs hysteresis of MOS transistors with FE in the gate stack or connected in series to the gate. Positive (CCW) hysteresis is dominated by FE P switching, while negative (CW) hysteresis is dominated by charge trapping-detrapping. Cancelation of these effects, for limited experimental conditions, leads to negligible hysteresis.

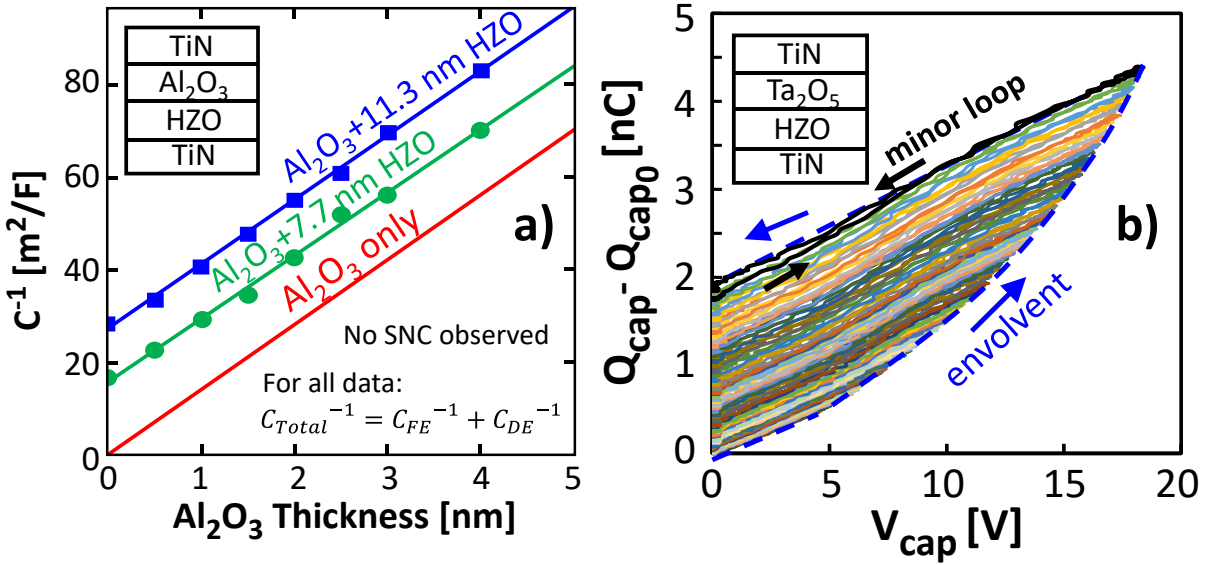

**Fig. 4**. a) Small signal capacitance-voltage measurements performed on DE-FE bilayer capacitors in ref [24], shows the capacitance of the stack having no anomalous behavior (i.e. capacitance of the stack is the conventional series capacitance) and no evidence of negative capacitance, thus ruling out stabilized negative capacitance models. b) Charge-voltage trajectory for a DE-FE bilayer capacitor from data in ref. [25], showing conventional FE hysteresis. The experiment starts with a large negative voltage ensuring all domains are switched in the same direction, and resulting in an initial capacitor charge $Q_{cap_0}$. A sequence of µs range pulses of increasing voltage is applied. As peak pulse voltage is increased, some domains in the FE start to switch polarization resulting in charge accumulation. Since the peak voltage increment (from the peak voltage of the previous pulse) for each pulse is small, only few domains switch in each pulse, but the effect is cumulative. Normal hysteretic behavior is observed for each pulse (minor loops) and the overall FE hysteresis (charge accumulation through the experiment) is depicted by the envolvent (blue dotted line).

**Supplemental Materials**

**Additional notes on transient negative capacitance and ferroelectric switching speed**

Although interesting physics are involved in TNC (or any mechanism for NC involving FE P switching), this phenomenon is not attractive for scaled low power electronics [28-32]. If a stabilized NC not requiring ferroelectric polarization switching was possible, it would also be present at high switching speeds, and would be lossless. In contrast, FE switching is too slow a process (e.g. nucleation and grow of domains during $P$ reversal) for high speed modern electronics and involves in general FE hysteresis losses. An effort has been made by several researchers to attempt to prove that FE polarization switching driven by electric fields can be fast enough to be compatible with transistor switching times in modern electronics (which would require FE switching in $ps$ timescales). Data for these publications consistently show that higher fields (voltages) can reduce switching time; however, at very high fields, it is well known that insulators break down. Even at fields (voltages) lower than the breakdown field, repeated pulses (cycling) induce breakdown at lower voltages. These effects limit the operation of devices, in practical applications, to fields that would ensure reasonable reliability. From the data shown so far (e.g. see Supplemental Materials Fig. 1, data from ref [48]), it is clear that switching speeds in the $ns$ range (still too slow for modern electronics) require very high voltages (10 V) for which device reliability is unacceptable (breakdown fields of dielectrics are typically of ~ 1 V/nm or lower [49]). Higher voltages would result in breakdown even before completing one operating cycle.

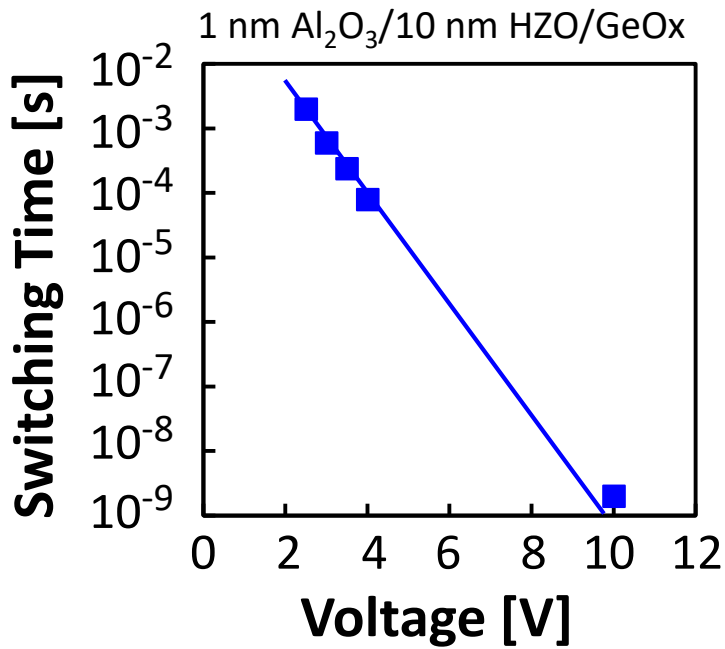

**Supplemental Materials Fig. 1.** Ferroelectric $P$ switching time vs voltage (proportional to electric field $E$), data from ref [48]. Note that breakdown fields are typically below 10 MV/cm (1 V/nm) [49]. Faster switching times are not accessible due to breakdown.

Measurements of characteristic ferroelectric switching times in the order of 220 ps were reported [50], with the data obtained by an ultrafast measurement technique using a fast rise electrical pulse, however, these switching speeds are still slow compared to requirements for high performance electronics. It is generally accepted that the timescales for polarization reversal in FE by the nucleation and growth of domains and induced by electrical pulses (switching in this case is an incoherent process) is limited to 100s of ps [51]. A coherent excitation by laser pulses can lead, however, to much higher switching speeds [51].

**Additional notes on physical foundations**

We assumed, following SNC models, that in Eq. 4 $\alpha_{FE} < 0$ and $\beta_{FE}$ or $\gamma_{FE} > 0$ for a ferroelectric material in the FE state. In general, the Landau model can describe a FE with a second order FE transition by taking $\alpha_{FE} < 0$ and $\beta_{FE} > 0$ in the FE state, and it suffices to use the expansion to the 4$^{th}$ power in polarization [36]. In contrast, FE with first order FE transitions can be described with $\beta_{FE} < 0$ and $\gamma_{FE} > 0$; in this case, there is a narrow temperature range for which the FE state can be achieved with $\alpha_{FE} > 0$ [37]. This case (FE state with $\alpha_{FE} > 0$) is not considered in models of SNC since according to the arguments of such models, it would not lead to SNC, so it is not discussed here.

It is important to note that Landau makes an effort to use different notation in order to distinguish different quantities and variables. He distinguishes in his notation between the free energies per unit volume and the free energies of the entire systems. Landau also distinguishes between the work (against electrostatic forces) done to create a free charge configuration in the presence of a material, $R_f = -W_f$, (here we use the sub-index "$f$", for free charges; $W_f$ is the work done by the electrostatic forces on the free charges) which only in systems without losses leads to the definition of an energy, $U_f = R_f = -W_f$; and the free energy of the material as a function of internal material configurational parameters (typically using the sub-index "$b$" for bound charges in the material) at fixed external conditions (both concepts get entangled when considering the electric field as independent variable). For a large planar capacitor of area $A$ and distance $d$ between its plates, the work $R_f$ to create the free charge configuration with total charge $Q_f$ in its plates is given by:

$$Ad\, r_f(\sigma_f) = R_f(Q_f) = \int V(Q_f)\, dQ_f \qquad (1)$$

Where $V$ is the voltage applied across the plates. The capacitance $C$ is given by:

$$C = \left(\frac{A}{d}\right)\left(\frac{\partial^2 r_f}{\partial \sigma_f^2}\right)^{-1} \qquad (2)$$

For a dielectric capacitor:

$$u_f^{DE}(\sigma_f) \equiv r_f^{DE}(\sigma_f) = \frac{\sigma_f^2}{2\epsilon} \qquad (3)$$

where $U_f^{DE}(\sigma_f) = Ad\, u_f^{DE}(\sigma_f)$ is the energy of the free charge configuration in the plates of the DE capacitor, and $\epsilon$ the permittivity of the dielectric. This expression is only valid for linear dielectrics. For ferroelectric materials undergoing typical hysteretic cycles, there is loss so that there is a net work put into the system in each cycle. Thus a single valued energy function cannot be defined; rather, the work put into the system (that would be defined as energy) keeps increasing with each cycle [32].

Landau also distinguishes between total electric field (macroscopic) at a point (here, $E$), external field (created by charges external to the material, here, $E_{ext}$) and displacement field (here, $D$).

Landau also distinguishes the different free energies of a material depending on the different possible constraints for their minimization (i.e. independent variables), e.g. using strain or stress as independent variables (here, at fixed electric field, for the free energy per unit volume of the material using strain or stress, we use $F_b(D, P, \varepsilon_{ij})$ and $\Phi_b(D, P, \sigma_{ij})$ respectively, where $P$ is the polarization per unit volume, $\varepsilon_{ij}$ the strain tensor and $\sigma_{ij}$ the stress tensor). Landau also distinguishes between the free energies with displacement field as an independent variable (e.g. $\Phi_b(D, P, \sigma_{ij})$ and those with electric field as independent variable, which he differentiates by the accent ~: e.g. $\widetilde{\Phi}_b(E, P, \sigma_{ij})$, etc. We also use free energies defined with external charges as independent variables (equivalently fixed external electric fields, which for many problems is equivalent to fixed displacement) and we notate these with an asterisk, e.g. as $\Phi_b^*(E_{ext}, P, \sigma_{ij})$.

In addition, we note that within the formalism from ref [42], using eqs. (6) and (7), all known materials (including ferroelectrics) would need to satisfy $\alpha' > 0$. Simply consider a large capacitor with a hypothetical homogenous material of thickness $d$ for which $\alpha' < 0$, since $D = -\sigma_f$ (where $\sigma_f$ is the free charge areal density in the top plate, and $D$ is the upward component of the displacement field), at small $\sigma_f$ values (which can be easily prepared experimentally), we have $E = -\alpha'\sigma_f$, i.e. $V = -d.E = \alpha'\sigma_f$, so that the material is an intrinsic negative capacitor (to our knowledge still not found in nature).

**Additional notes on models of stabilization of a FE in a non-switching negative capacitance state by the presence of an adjacent dielectric**

In ref [23], the free energies $\widetilde{\Phi}_b(E, P)$ are used for the FE and DE layers (which only leads to correct equations of state when minimized under constant electric fields). A condition of fixed total voltage at the capacitor plates for a bilayer, or fixed total voltage from the top of one capacitor to the bottom of the other capacitor, is used as constraint, and the addition of the free energies of the DE and FE layers ($\widetilde{\Phi}_b^{FE}(E^{FE}, P^{FE}) + \widetilde{\Phi}_b^{DE}(E^{DE}, P^{DE})$) is minimized under this constraint. It is clear that this condition is not equivalent to fixed fields $E^{FE}$ and $E^{DE}$ in each layer. We consider, to illustrate this point, another example of exactly the same mistake. We start from the free energy $\widetilde{\Phi}_b(E, P)$ of a FE in a capacitor, as given by eq. (4), with $\alpha < 0$, as explained above, leading to the FE behavior. Let's assume the free charge density in the plates of the capacitor is 0: $\sigma_f = 0$. The field in the FE is then $E = -\frac{(\sigma_f + P)}{\epsilon_0} = -\frac{P}{\epsilon_0}$. Replacing this value of the field in eq. (4), we obtain:

$$\widetilde{\Phi}_b(E, P) = \frac{\alpha}{2}P^2 + \frac{\beta}{4}P^4 + \frac{\gamma}{6}P^6 + \frac{P^2}{\epsilon_0} \quad (8)$$

$$\Rightarrow \widetilde{\Phi}_b(E, P) = \left(\frac{\alpha}{2} + \frac{1}{\epsilon_0}\right)P^2 + \frac{\beta}{4}P^4 + \frac{\gamma}{6}P^6 \quad (9)$$

Under the condition $-\frac{\alpha}{2} = \frac{|\alpha|}{2} < \frac{1}{\epsilon_0}$, the free energy given by eq. (4) has a minimum at $P = 0$ (rather than a maximum as expected for FE with $\alpha < 0$). Has the FE suddenly decided to behave as a dielectric because of this mathematical trick? The answer is obviously *no*. The source of this invalid argument is indeed that the minima of $\widetilde{\Phi}_b(E, P)$ at fixed external charges or external fields (rather than fixed electric field) has no physical meaning.

**Additional notes on experimental evidence**

Some experiments using epitaxial perovskite DE-FE structures have shown an enhancement of capacitance of the stack compared to the capacitance of the DE layer alone [3]. For coupled systems, there is no reason why the system properties should be a simple combination of the properties of the layers. In a rather crude analogy, we consider high-$T_c$ superconductors in the superconducting state: while the resistance along the stacking direction is 0, we cannot conclude that some layers (e.g. Cu-O planes) have a negative resistance simply because other layers (e.g. Y-O or Ba-O layers) have a positive resistance when isolated. The properties of coupled systems are not a simple combination of the properties of each component.